\documentstyle[12pt]{article}
\begin{document}
\begin{center}
{\bf{Comment on the paper by Arik, G\"{u}n and Yildiz (hep-th/0212129) on "Invariance Quantum Group of the 
Fermionic Oscillator"}}

\vspace{0.5cm}

R.Parthasarathy{\footnote{Permanent address: The Institute of Mathematical Sciences, Chennai 600 113, India: 
e-mail: sarathy@imsc.ernet.in}} \\
 and \\ 
K.S.Viswanathan{\footnote{e-mail address: kviswana@sfu.ca}} 

Department of Physics, Simon Fraser University \\
Burnaby, B.C., Canada V5A 1S6.
\end{center}

\vspace{0.5cm}

\begin{center}
{\it{Abstract}}
\end{center}

A non-trivial q-deformation of the fermionic oscillator proposed by us in 1991 is recalled in view of the result 
of the paper cited in the title.

\vspace{0.5cm}

In the paper by Arik, G\"{u}n and Yildiz [1], the authors considered the fermionic oscillator 
\begin{eqnarray}
cc^*+c^*c&=&1, \nonumber \\
c^2&=& 0.
\end{eqnarray}
and state that the above algebra (1) does not admit a q-deformation. Then, 
they have shown that the structure of the inhomogeneous invariance group of fermionic oscillator is a quantum group.

\vspace{0.5cm}

The fact that when $c^2=0$, the fermionic oscillator algebra does not admit a q-deformation has been pointed out in 
1991 by Jing and Xu [2]. Considering {\it{a possible}} q-deformation of the fermion oscillator algebra proposed 
by Chaichian and Kullish [3], namely,
\begin{eqnarray}
c_qc_q^*+qc_q^*c_q= q^N, \nonumber \\
c_q^2=0, \ {[N,c_q^*]}=c_q^*, {[N,c_q]}=-c_q, 
\end{eqnarray}
and making a transformation $c= q^{-\frac{N}{2}}c_q\ ,\ c^*=c^*_q q^{-\frac{N}{2}}$, the above algebra (2) reduces to 
(1), showing that (1) does not admit a q-deformation such as in (2).

\vspace{0.5cm}

However, we [4] in 1991, have proposed {\it{a non-trivial}} q-deformation of the fermion oscillator algebra (1), as 
\begin{eqnarray}
f_qf_q{\dagger}+\sqrt{q}f_q^{\dagger}f_q = q^{-\frac{N}{2}}, \nonumber \\
{[N,f_q^{\dagger}]}=f_q^{\dagger},\ {[N,f_q]}=-f_q;\ f_q^2\neq 0, 
\end{eqnarray}
which cannot be reduced to (1) and hence non-trivial. In fact, we have pointed out that in q-deforming the 
fermionic oscillator algebra, the Pauli principle has to be relaxed. {\it{The relation $f_q^2\neq 0$, allows more than two 
q-fermions in a given quantum state.}} The Fock space of such q-fermion algebra has been constructed in [4] using 
fermionic q-numbers,
\begin{eqnarray}
{[n]}^F_q&=& \frac{ q^{-\frac{n}{2}}-(-1)^nq^{\frac{n}{2}}}{q^{-\frac{1}{2}}+q^{\frac{1}{2}}},
\end{eqnarray}
and in the limit $q\rightarrow 1$, the Fock space reduces to vacuum and one-fermion state, thereby recovering the 
Pauli principle in this limit. So it is possible to q-deform the fermionic oscillator algebra (1) and such a 
q-deformed algebra is (3).

\vspace{0.5cm}

Further, using (7), (8) and (28) of [1], it follows that the action of their $A$ matrix in (5), produces same 
algebra as in (2) (upto overall normalization) {\it{but}} with $(c')^2\neq 0$, whose structure is different from (3).

\vspace{0.5cm}

{\noindent{\bf{References.}}}

\begin{enumerate}
\item M.Arik, S.G\"{u}n and A.Yildiz, {\it{Invariance quantum group of fermionic oscillator}}, hep-th/02121219.
\item S.Jing and J.Xu, J.Phys.A: Math.Gen. {\bf{24}} (1991) L891.
\item M.Chaichian and P.I.Kullish, CERN Preprint, Th.5969/90, November 1990.
\item R.Parthasarathy and K.S.Viswanathan, J.Phys.A: Math.Gen. {\bf{24}} (1991) 613.
\end{enumerate}
\end{document}